# Innovations orthogonalization: a solution to the major pitfalls of EEG/MEG "leakage correction"

Roberto D. Pascual-Marqui[1,2], Rolando J. Biscay[3], Jorge Bosch-Bayard[4], Pascal Faber[1], Toshihiko Kinoshita[2], Kieko Kochi[1], Patricia Milz[1], Keiichiro Nishida[2], Masafumi Yoshimura[2]

1: The KEY Institute for Brain-Mind Research, Department of Psychiatry, Psychotherapy, and Psychosomatics, University Hospital of Psychiatry, Zurich, Switzerland
2: Department of Neuropsychiatry, Kansai Medical University, Osaka, Japan
3: Centro de Investigación en Matemáticas. A.C., Jalisco S/N, Col. Valenciana, Guanajuato, 36023, México
4: Departamento de Neurobiología Conductual y Cognitiva, Instituto de Neurobiología, Universidad Nacional Autónoma de México. Boulevard Juriquilla 3001, Querétaro, 76230, México

Corresponding author: RD Pascual-Marqui
pascualmarqui@key.uzh.ch ; www.uzh.ch/keyinst/loreta.htm
scholar.google.com/citations?user=pascualmarqui

## 1. Abstract

The problem of interest here is the study of brain functional and effective connectivity based on non-invasive EEG-MEG inverse solution time series. These signals generally have low spatial resolution, such that an estimated signal at any one site is an instantaneous linear mixture of the true, actual, unobserved signals across all cortical sites. False connectivity can result from analysis of these low-resolution signals. Recent efforts toward "unmixing" have been developed, under the name of "leakage correction". One recent noteworthy approach is that by Colclough et al (2015 NeuroImage, 117:439-448), which forces the inverse solution signals to have zero cross-correlation at lag zero. One goal is to show that Colclough's method produces false human connectomes under very broad conditions. The second major goal is to develop a new solution, that appropriately "unmixes" the inverse solution signals, based on innovations orthogonalization. The new method first fits a multivariate autoregression to the inverse solution signals, giving the mixed innovations. Second, the mixed innovations are orthogonalized. Third, the mixed and orthogonalized innovations allow the estimation of the "unmixing" matrix, which is then finally used to "unmix" the inverse solution signals. It is shown that under very broad conditions, the new method produces proper human connectomes, even when the signals are not generated by an autoregressive model.

## 2. Introduction

The problem of interest here is the study of brain connectivity based on non-invasive measurements of scalp electric potential differences (EEG) and extracranial magnetic fields (MEG). These measurements can be used to estimate time series of cortical electric neuronal activity, which in turn can be used for estimating functional and effective connectivity (see e.g. Valdes-Sosa et al 2011, Friston 2011).

However, these estimators of cortical activity are characterized by low spatial resolution. In the case of estimators based on distributed, discrete, linear inverse solutions, the resolution is characterized by the Backus and Gilbert resolution matrix (Backus and Gilbert 1968). Based on the properties of the resolution matrix, it can be shown that the inverse solution signal at any one cortical site corresponds to an instantaneous linear mixture of the true, unobserved, actual signals all over the cortex.

Thus, connectivity inference based on inverse solution signals can be false and erroneous.

An early particular solution to this problem was given by Nolte et al (2004). It is based on the idea that the instantaneous mixture (due to low resolution and to volume conduction), will affect mainly the real part of the complex valued coherence while the imaginary part of the coherence is mainly determined by lagged, non-instantaneous connections.

This early work by Nolte et al (2004) established a simple but important principle for making connectivity inference: the instantaneous, lag zero connectivity between time series is strongly affected by low resolution and volume conduction.

More recently, and related to this principle, it seems that the following reasoning has been established in some of the literature:
"If low-resolution and volume-conduction increase instantaneous cross-correlation, then forcing the signals to have zero instantaneous cross-correlation will make them adequate for correct connectivity inference."

Three exemplary papers that advocate the use of signal orthogonalization methods for achieving this goal, denoted by their authors as "leakage correction" procedures, are Brookes et al 2012, Hipp et al 2012, and Colclough et al 2015.

The leakage correction algorithms of Brookes et al 2012 and of Hipp et al 2012 are defined for treating one pair of signals at a time. In contrast, the more recent leakage correction method of Colclough et al 2015 is multivariate.

Here we contend that actual signals of cortical electric neuronal activity are not orthogonal, i.e. they are not instantaneously uncorrelated with zero cross-correlation at lag zero. We further contend that "leakage correction" applied to inverse solution signals is prone to producing false human connectomes. And going one step further, we additionally contend that even in the case of non-mixed signals, "leakage correction" will very likely produce false human connectomes.

One goal of this paper is to present evidence to support these assertions. It is essential to bring awareness to the connectome research community of these serious pitfalls of the leakage correction method applied to electrophysiological signals.

The central goal of this paper is to present a solution to this problem. The main assumption here is that the multivariate autoregressive innovations are orthogonal, i.e. uncorrelated with zero cross-correlation at lag zero, which is one of the main requirements of the "innovation approach" in time series analysis, as detailed in Ozaki 2012. This will allow the estimation of a mixing matrix, which is then inverted and used for resolving, i.e "unmixing" the measured (or estimated) signals. We show that "innovations orthogonalization" provides adequate connectivity estimates even when the actual signals are not generated by an autoregression (AR).

## 3.     The resolution of a mixture of time series by innovations orthogonalization

Let $\mathbf{X}_t \in \mathbb{R}^{p \times 1}$ denote "$p$" unobserved, zero mean, time series, with $t = 1 \ldots N_T$ denoting the time sample. Consider the AR of order "$q$" for the unobserved $\mathbf{X}_t$:

Eq. 1 $\quad \mathbf{X}_t = \sum_{k=1}^{q} \mathbf{A}_k \mathbf{X}_{t-k} + \boldsymbol{\varepsilon}_t$

with AR coefficients $\mathbf{A}_k \in \mathbb{R}^{p \times p}$, and with independent innovations $\boldsymbol{\varepsilon}_t \in \mathbb{R}^{p \times 1}$ having a diagonal covariance matrix $\mathbf{S}_{\varepsilon\varepsilon}$, i.e.:

Eq. 2 $\quad \mathbf{S}_{\varepsilon\varepsilon} = diag(\mathbf{S}_{\varepsilon\varepsilon}) = \frac{1}{(N_T - p)} \boldsymbol{\varepsilon}^T \boldsymbol{\varepsilon} = \frac{1}{(N_T - p)} \sum_{t=p+1}^{N_T} \boldsymbol{\varepsilon}_t \boldsymbol{\varepsilon}_t^T$

where the superscript "$T$" denotes vector-matrix transposition, the operator $diag(\bullet)$ returns a matrix with off-diagonal elements set to zero, and where $\boldsymbol{\varepsilon} \in \mathbb{R}^{(N_T - p) \times p}$ denotes the time series written in matrix notation including all available time samples.

The requirement of independence of the innovations is one of the main properties of the "innovation approach", as detailed in, e.g. Ozaki 2012.

Note that the independent innovations can be written uniquely as:

Eq. 3 $\quad \boldsymbol{\varepsilon} = \mathbf{V}\mathbf{D}$

with orthonormal $\mathbf{V} \in \mathbb{R}^{(N_T - p) \times p}$:

Eq. 4 $\quad \mathbf{V}^T \mathbf{V} = \mathbf{I}$

and diagonal matrix $\mathbf{D} \in \mathbb{R}^{p \times p}$:

Eq. 5 $\quad \mathbf{D} = \left[ (N_T - p) \mathbf{S}_{\varepsilon\varepsilon} \right]^{1/2}$

The uniqueness of $\mathbf{D}$ is given by the fact that $\mathbf{S}_{\varepsilon\varepsilon}$ is a diagonal matrix, with unique square root. Furthermore:

Eq. 6 $\quad \mathbf{V} = \boldsymbol{\varepsilon} \left[ (N_T - p) \mathbf{S}_{\varepsilon\varepsilon} \right]^{-1/2}$

It will be assumed that the estimated inverse solution signals, denoted as $\mathbf{Y}_t \in \mathbb{R}^{p \times 1}$, correspond to an instantaneous linear mixture of the unobserved time series:

Eq. 7 $\quad \mathbf{Y}_t = \mathbf{M}\mathbf{X}_t$

where the symmetric, non-singular mixing matrix $\mathbf{M} \in \mathbb{R}^{p \times p}$ has diagonal elements equal to 1, and absolute values of off-diagonal elements smaller than 1:

Eq. 8 $\quad \begin{cases} [\mathbf{M}]_{ij} = [\mathbf{M}]_{ji} \\ [\mathbf{M}]_{ii} = 1 \\ \left| [\mathbf{M}]_{ij} \right| < 1 \; , \; \forall i \neq j \end{cases}$

In the ideal case, there is no mixture, and $\mathbf{M}$ is the identity matrix $\mathbf{I}$. This means that each observed univariate time series $[\mathbf{Y}_t]_i$, with $i = 1 \ldots p$, is identical to its corresponding unobserved univariate time series $[\mathbf{X}_t]_i$.

In a "near ideal" case, the mixing matrix has ones on the diagonal, and close to zero off-diagonal values. This means that each observed univariate time series $[\mathbf{Y}_t]_i$, with $i = 1 \ldots p$, is close in value to its corresponding unobserved univariate time series $\left[\mathbf{X}_t\right]_i$, except for a weak instantaneous linear mixture of all other unobserved univariate time series $\left[\mathbf{X}_t\right]_j$, $\forall j \neq i$.

The AR for the observations $\mathbf{Y}_t$ is:

Eq. 9 $\quad \mathbf{Y}_t = \sum_{k=1}^{q} \mathbf{B}_k \mathbf{Y}_{t-k} + \mathbf{\eta}_t$

with:

Eq. 10 $\quad \mathbf{B}_k = \mathbf{M}\mathbf{A}_k\mathbf{M}^{-1}$

and:

Eq. 11 $\quad \mathbf{\eta}_t = \mathbf{M}\mathbf{\varepsilon}_t$

and where the mixed innovations $\mathbf{\eta}_t$ are now correlated:

Eq. 12 $\quad \mathbf{S}_{\eta\eta} = \mathbf{M}\mathbf{S}_{\varepsilon\varepsilon}\mathbf{M}$

The mixed time series $\mathbf{Y}_t$ can be resolved to obtain $\mathbf{X}_t$ as follows.

First, given the observations $\mathbf{Y}_t$, fit the AR model of order "$q$" specified in Eq. 9. As usual, the order "$q$" can be estimated using Akaike's information criterion (AIC) (Akaike 1974).

This first step gives the mixed (correlated) innovations $\mathbf{\eta}_t \in \mathbb{R}^{p \times 1}$, which in matrix notation, including all available time samples, is $\mathbf{\eta} \in \mathbb{R}^{(N_T - p) \times p}$.

Second, the matrix $\mathbf{\varepsilon} \in \mathbb{R}^{(N_T - p) \times p}$ of independent, uncorrelated (i.e. orthogonal) innovations, expressed as in Eq. 3 and Eq. 11, is estimated by solving the following problem:

Eq. 13 $\quad \min_{\mathbf{V},\mathbf{D}} tr\left[ (\mathbf{\eta} - \mathbf{VD})^T (\mathbf{\eta} - \mathbf{VD}) \right]$, under constraints: $\mathbf{V}^T\mathbf{V} = \mathbf{I}$, $\mathbf{D} = diag(\mathbf{D})$

where the operator $tr(\bullet)$ returns the trace of a matrix.

The problem in Eq. 13 corresponds to the "Orthogonal Procrustes" problem, stated and solved algorithmically by Everson 1999. The algorithm as implemented in this present study is given in the Appendix 1. We acknowledge here that we were made aware of this algorithm by the work of Colclough et al 2015, who applied this algorithm to the inverse solution signals $\mathbf{Y}_t$ instead of innovations. [The method of Colclough et al 2015 for signal orthogonalization (leakage correction) is briefly outlined below, in the section entitled "4. The leakage correction method of Colclough et al 2015".]

It is very important to emphasize a special property of Everson's orthogonalization: the orthogonal signals are unique, i.e. they are in a one-to-one correspondence with the original signals. This is distinct from principle and independent components analysis (PCA and ICA), where there are sign and permutation indeterminacies.

The solution to the problem in Eq. 13 is denoted as $\mathbf{V}_{io}$ and $\mathbf{D}_{io}$, which gives the following estimator for the orthogonal innovations of the unobserved time series:

Eq. 14 $\quad \mathbf{\varepsilon}_{io} = \mathbf{V}_{io}\mathbf{D}_{io}$

where "$io$" as subscript and superscript corresponds to the method of *i*nnovations *o*rthogonalization.

Third, the mixing matrix is now readily computed as the least squares estimator from Eq. 11 for all available time samples:

Eq. 15 $\quad \mathbf{M} = \left( \boldsymbol{\varepsilon}_{io}^T \boldsymbol{\varepsilon}_{io} \right)^{-1} \boldsymbol{\varepsilon}_{io}^T \boldsymbol{\eta} = \mathbf{D}_{io}^{-1} \mathbf{V}_{io}^T \boldsymbol{\eta}$

Finally, the "*unmixing*" matrix $\mathbf{M}^{-1}$ is used for resolving the mixture into the "unmixed" time series:

Eq. 16 $\quad \hat{\mathbf{X}}_t^{io} = \mathbf{M}^{-1} \mathbf{Y}_t$

or:

Eq. 17 $\quad \hat{\mathbf{X}}^{io} = \mathbf{Y} \mathbf{M}^{-1}$

where $\mathbf{Y} \in \mathbb{R}^{N_T \times p}$ denotes the observed time series written in matrix notation, including all available time samples.

The estimator for the unmixed time series (Eq. 16, Eq. 17) is now used for connectivity analysis.

## 4. The leakage correction method of Colclough et al 2015

In this case, the basic assumption is that the actual unobserved time series $\mathbf{X}_t \in \mathbb{R}^{p \times 1}$ are orthogonal, i.e. uncorrelated, with zero cross-correlations at lag zero:

Eq. 18 $\quad \mathbf{S}_{xx} = diag(\mathbf{S}_{xx}) = \frac{1}{N_T} \mathbf{X}^T \mathbf{X} = \frac{1}{N_T} \sum_{t=1}^{N_T} \mathbf{X}_t \mathbf{X}_t^T$

where $\mathbf{X} \in \mathbb{R}^{N_T \times p}$ denotes the time series written in matrix notation, including all available time samples.

In what follows, "*lc*" as subscript and superscript is used to clarify that the method used is that of "*l*eakage *c*orrection", defined as signal orthogonalization. This is not to be confused with innovations orthogonalization "*io*".

The orthogonality assumption implies that unobserved signals $\mathbf{X}$ can be written uniquely as:

Eq. 19 $\quad \mathbf{X} = \mathbf{V}_{lc} \mathbf{D}_{lc}$

with orthonormal $\mathbf{V}_{lc} \in \mathbb{R}^{N_T \times p}$:

Eq. 20 $\quad \mathbf{V}_{lc}^T \mathbf{V}_{lc} = \mathbf{I}$

and diagonal matrix $\mathbf{D}_{lc} \in \mathbb{R}^{p \times p}$:

Eq. 21 $\quad \mathbf{D}_{lc} = \left[ N_T \mathbf{S}_{xx} \right]^{1/2}$

The uniqueness of $\mathbf{D}_{lc}$ is given by the fact that $\mathbf{S}_{xx}$ is assumed to be a diagonal matrix, with unique square root. Furthermore:

Eq. 22 $\quad \mathbf{V}_{lc} = \mathbf{X} \left[ N_T \mathbf{S}_{xx} \right]^{-1/2}$

In the work of Colclough et al 2015, the observed time series $\mathbf{Y}_t \in \mathbb{R}^{p \times 1}$ are assumed to be a mixture (due to "leakage") of the orthogonal unobserved time series $\mathbf{X}_t \in \mathbb{R}^{p \times 1}$. And the "leakage corrected" estimator for the unobserved time series is obtained by a direct application of Everson's algorithm (see the Appendix 1) for solving the orthogonal Procrustes problem to the measurements:

Eq. 23 $\quad \min_{\mathbf{V}_{lc}, \mathbf{D}_{lc}} tr \left[ \left( \mathbf{Y} - \mathbf{V}_{lc} \mathbf{D}_{lc} \right)^T \left( \mathbf{Y} - \mathbf{V}_{lc} \mathbf{D}_{lc} \right) \right]$ , under constraints: $\mathbf{V}_{lc}^T \mathbf{V}_{lc} = \mathbf{I}$ , $\mathbf{D}_{lc} = diag(\mathbf{D}_{lc})$

which gives $\left(\mathbf{V}_{lc}, \mathbf{D}_{lc}\right)$, and thus:

Eq. 24  $\hat{\mathbf{X}}^{lc} = \mathbf{V}_{lc} \mathbf{D}_{lc}$

The estimator for the unmixed time series (Eq. 24) is now used for connectivity analysis.

## 5. False human connectomes ensue from leakage correction, i.e. signal orthogonalization, under very broad conditions, even without mixing

Here we contend that under very broad conditions, "leakage correction", i.e. signal orthogonalization, of time series of cortical electric neuronal activity will lead to false connectivities.

Compelling evidence is based on the fact that under a wide range of conditions, electrophysiological signals obey a multivariate autoregressive model (see e.g. Valdes-Sosa et al 2011, Bressler and Seth 2011), which causally introduces non-zero values of cross-correlation at lag zero.

This means that even if the signals are unmixed (i.e. without leakage), there is significant lag zero cross-correlation.

To see this intuitively, consider the case of an autoregression for three or more signals. If two or more signals receive causal influence with common lag from another signal, then the receiving signals can have significant instantaneous zero-lag cross-correlation. Leakage correction will then significantly distort the true connectivity pattern. A toy network (copied from Pascual-Marqui et al 2014) with these properties is shown in Figure 1. The full specification of the causal autoregressive model is shown in the Appendix 2.

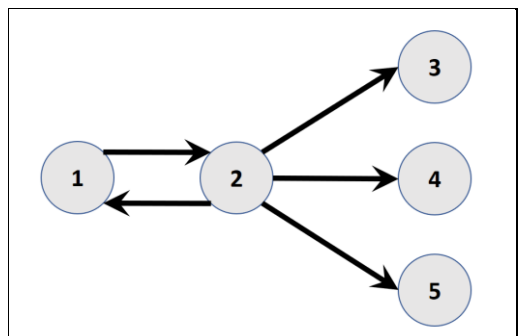

Figure 1: Toy example of five interconnected nodes (copied from Pascual-Marqui et al 2014). The full specification of the model is shown in the Appendix 2. The number of time samples used in the toy example is 25600.

By definition and construction, this toy example is lacking any form of lag-zero connection. Furthermore, there is no "mixing" here, there is no "leakage". But this is unknown to the researcher that wants to study the connections. The researcher calculates the cross-correlation matrix at zero-lag, which is significant (number of time samples 25600) and shown in Table 1.

Table 1: Cross-correlation matrix at zero-lag for the toy example in Figure 1 and Appendix 2. The number of time samples used in the toy example is 25600.

|  |  |  |  |  |
|---|---|---|---|---|
| 1 | −0.83132 | −0.40895 | −0.41049 | −0.40963 |
| −0.83132 | 1 | 0.396376 | 0.398864 | 0.398153 |
| -0.40895 | 0.396376 | 1 | 0.872185 | 0.87103 |
| -0.41049 | 0.398864 | 0.872185 | 1 | 0.869128 |
| -0.40963 | 0.398153 | 0.87103 | 0.869128 | 1 |

Based on the observed significant cross-correlations at lag-zero, the researcher applies the multivariate "leakage-correction" of Colclough et al 2015 to the signals.

The isolated effective coherence (iCoh) as a function of frequency (Pascual-Marqui et al 2014) for the actual signals and for the "leakage corrected" signals are shown in Figure 2.

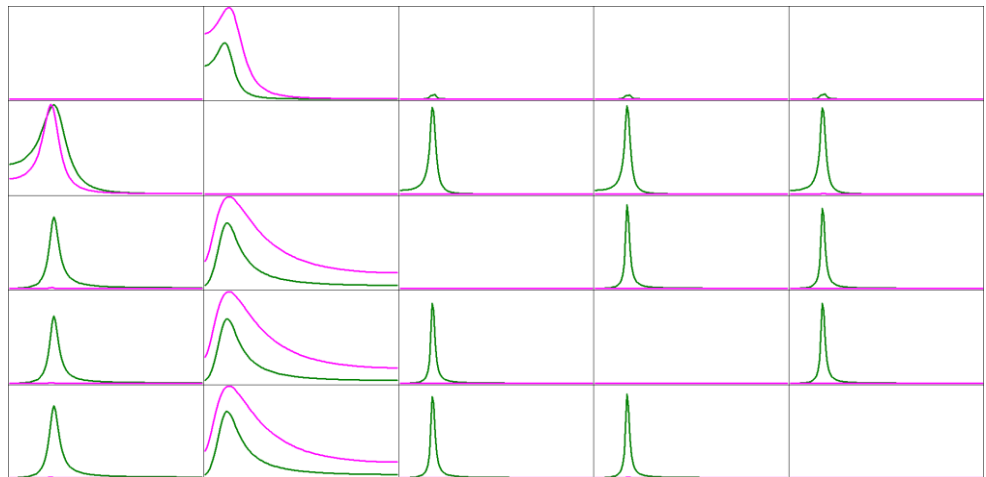

Figure 2: Isolated effective coherence (iCoh) as a function of frequency, for the toy example in Figure 1 and Appendix 2. The number of time samples used in the toy example is 25600. Horizontal frequency axis range 1-127Hz. Vertical iCoh axis range 0-1. A column is a causal sender, a row is a causal receiver. Magenta curves are correct values, green curves are for the "leakage-corrected" signal.

In Figure 2, a column is a causal sender, and a row is a causal receiver. Magenta curves are correct values, while green curves correspond to the "leakage-corrected" signals. Note the magenta curves in column 1, correctly indicating that node 1 is a causal sender to node 2 only. Also correct for the magenta curves are the connections from node 2 to nodes 1, 3, 4, and 5. There are no other correct connections. Now note the incorrect peaks in the leakage-corrected case (green curves) in column 1, falsely indicating that node 1 sends to nodes 3, 4, and 5. More worrisome is the fact that the graph is full of very high false connections for almost all pairs of nodes for the leakage-corrected signal (green curves).

Two very important conclusions follow:
1. Non-zero values of cross-correlation at lag-zero can occur without "mixture", without "leakage".
2. If "leakage correction" is applied to signals that are falsely thought to be "mixed", then the estimated connectivities can be false and significantly different from the true connectivities.

The previous toy example corresponds to signals that are not mixed. If these signals are mixed, with an instantaneous mixing matrix with ones on the diagonal and all off-diagonal elements equal to 0.7, the "leakage correction" algorithm applied to the new mixed signals produces again false connections as shown in the Appendix 3.

## 6. False human connectomes ensue from leakage correction, i.e. signal orthogonalization, of signals of electric neuronal activity in oscillatory models, even without mixing

Consider three time series generated as noisy oscillations:

Eq. 25
$$\mathbf{X}_t = \begin{pmatrix} a_1 \sin(\omega_1(t-\tau_1)) + \varepsilon_{1,t} \\ a_2 \sin(\omega_2(t-\tau_2)) + \varepsilon_{2,t} \\ a_3 \sin(\omega_3(t-\tau_3)) + \varepsilon_{3,t} \end{pmatrix}$$

In the toy example constructed here, 100 epochs with 256 time samples each were generated. Sampling rate was 256 Hz. The fixed parameters were:

Eq. 26  $\omega_1 = \omega_2 = \dfrac{2\pi \times 10}{256}$ ; $\omega_3 = \dfrac{2\pi \times 17}{256}$ ; $\tau_1 = 0$ , $\tau_2 = -1$ , $\tau_3 = -2$

The noise $\varepsilon$ was independent uniformly distributed in the interval $\pm 0.9$. The amplitudes "$a$" changed from epoch to epoch, generated as independent uniformly distributed variables in the interval +0.5 to +1.5. The code is given in Appendix 4.

Signal 2 is a noisy delayed version of signal 1, with 10 Hz as main frequency. Signal 3 is an orthogonal oscillator at 17 Hz.

Note that by definition and construction, there is no instantaneous zero-lag connection between signals 1 and 2. By definition, there is a simple delay. However, the zero-lag cross correlation between signals 1 and 2 is 0.609, which is significant. Thus, a researcher might be tempted to perform leakage correction before estimating connectivities. Figure 3 displays the squared modulus of the coherences, for the original unmixed signals (magenta curves), and for the "leakage-corrected" signals (green curves).

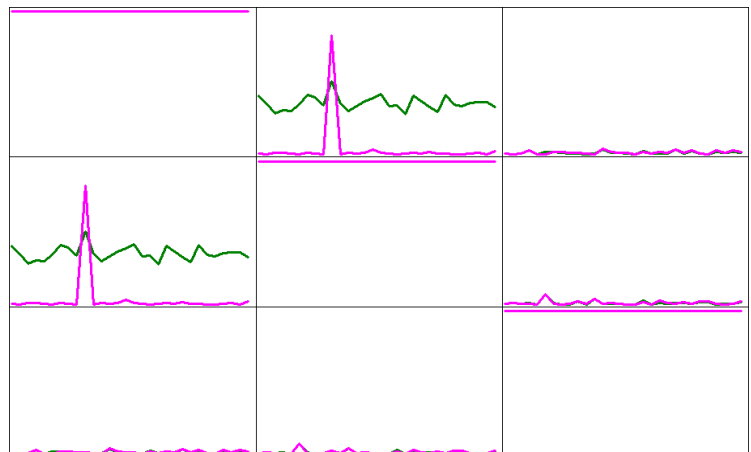

Figure 3: Squared modulus of coherence for three signals generated as in Eq. 25 and Eq. 26 (code is given in Appendix 4). Vertical coherence axis range 0-1. Horizontal frequency axis range 1-30 Hz. Magenta curves: true coherence for original three signals, displaying coherence peak at 10 Hz between signals 1 and 2. Green curves: false coherence between signals 1 and 2 introduced by applying leakage-correction.

In Figure 3, referring to signals 1 and 2, note the correct coherence peak at 10 Hz (magenta curves), while leakage correction (green curves) displays an incorrect frequency profile with false high coherence at all frequencies in the range 1-30 Hz.

We conclude again:
1. Non-zero values of cross-correlation at lag-zero can occur without "mixture", without "leakage".
2. If "leakage correction" is applied to signals that are falsely thought to be "mixed" due to "leakage", then the estimated connectivities can be false and significantly different from the true connectivities.

## 7. False human connectomes ensue from leakage correction, i.e. signal orthogonalization, in amplitude-amplitude coupling analysis

Early studies of amplitude-amplitude coupling can be found in Friston (1997), and Bruns et al 2000 and 2004. The paper by Colclough et al 2015 also studies amplitude-amplitude coupling, after performing leakage correction on inverse solution signals obtained from MEG recordings.

In this section, we use the same method as in van Wijk et al 2015 to generate toy examples of signals appropriate for the study of amplitude-amplitude coupling. The three signals in this case are generated as:

Eq. 27
$$\mathbf{X}_t = \begin{pmatrix} \{[1+0.5\times\sin(\omega_{s1}(t-\tau_{s1}))]\times\varepsilon_{1,t}\}\times\sin(\omega_{f1}(t-\tau_{f1})) \\ \{[1+0.5\times\sin(\omega_{s2}(t-\tau_{s2}))]\times\varepsilon_{2,t}\}\times\sin(\omega_{f2}(t-\tau_{f2})) \\ \{[1+0.5\times\sin(\omega_{s3}(t-\tau_{s3}))]\times\varepsilon_{3,t}\}\times\sin(\omega_{f3}(t-\tau_{f3})) \end{pmatrix}$$

This toy example has 25600 time samples, with a sampling rate of 256 Hz. The fixed parameters were:

Eq. 28  $\omega_{s1} = \frac{2\pi\times 2}{256}$ ; $\omega_{s2} = \frac{2\pi\times 3}{256}$ ; $\omega_{s3} = \frac{2\pi\times 5}{256}$ ; $\tau_{s1}=0$ ; $\tau_{s2}=\tau_{s3}=-4$

Eq. 29  $\omega_{f1} = \omega_{f2} = \frac{2\pi\times 22}{256}$ ; $\omega_{f3} = \frac{2\pi\times 28}{256}$ ; $\tau_{f1}=0$ ; $\tau_{f2}=-1$ ; $\tau_{f3}=-2$

The noise "ε" was independent, uniformly distributed in the interval +0.8 to +1.2. The code is given in Appendix 5.

Signal 1 oscillates at 22 Hz, and its amplitude is noisily modulated at 2 Hz; signal 2 oscillates at 22 Hz with a delay of 1 time sample relative to signal 1, and its amplitude is noisily modulated at 3 Hz; signal 3 oscillates at 28 Hz, and its amplitude is noisily modulated at 5 Hz.

By construction, this toy example has zero amplitude-amplitude coupling, i.e., the correlations between the instantaneous amplitudes (i.e. envelopes) of the signals are all zero.

Figure 4 shows four seconds of these signals.

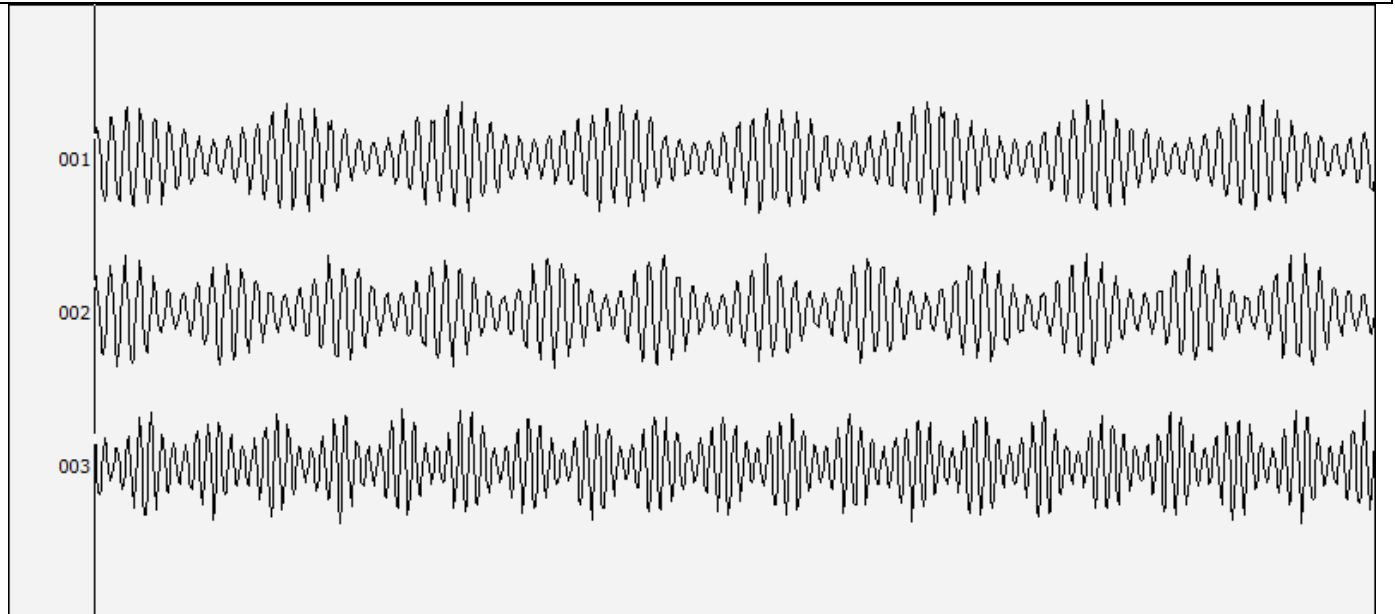

Figure 4: Four second segment of three signals used for the analysis of amplitude-amplitude coupling. See Eq. 27, Eq. 28, and Eq. 29 for their definitions, and see Appendix 5 for the program code used for their generation. These signals, by construction, have zero amplitude-amplitude coupling.

Despite the fact that all correlations between the instantaneous amplitudes are zero, the zero-lag cross correlation between signals 1 and 2 is +0.75, which is significant (sample size 25600). Thus, a researcher might be tempted to perform leakage correction before estimating the correlations between the envelopes of the signals. After leakage correction, the instantaneous amplitudes (estimated via the use of the discrete Hilbert transform) between signals 1 and 2 have a significant correlation coefficient of -0.40, which is false.

Thus, we state once more:
1. Non-zero values of cross-correlation at lag-zero can occur without "mixture", without "leakage".
2. If "leakage correction" is applied to signals that are falsely thought to be "mixed" due to "leakage", then the estimated connectivities can be false and significantly different from the true connectivities.

This example is particularly relevant in relation to the paper by Colclough et al 2015, in which amplitude-amplitude coupling was studied after applying leakage correction.

## 8. False human connectomes ensue from leakage correction, i.e. signal orthogonalization, of inverse solution signals

In this experiment, toy EEG recordings were generated as in Pascual-Marqui et al 2014. For the sake of completeness, some details are given here now.

The five time series generated as specified in Appendix 2, corresponding to the connections shown in Figure 1, were used as the time varying electric neuronal activities at the cortical locations specified in Table 2.

Table 2: Locations of cortical grey matter voxel used for generating EEG recordings.

| Region | Brodmann Area | MNI X mm | MNI Y mm | MNI Z mm |
|---|---|---|---|---|
| Superior frontal gyrus (left) | 10 | -25 | 65 | -5 |
| Middle Occipital Gyrus (right) | 18 | 20 | -100 | 5 |
| Postcentral Gyrus (left) | 3 | -50 | -20 | 60 |
| Middle Temporal Gyrus (left) | 21 | -65 | -15 | -15 |
| Middle Temporal Gyrus (right) | 21 | 70 | -20 | -10 |

Figure 5 illustrates the five cortical regions and their effective causal connections.

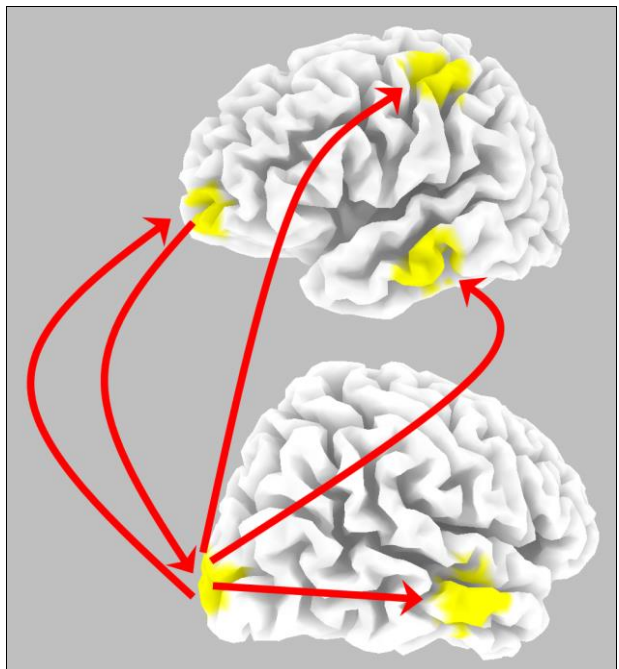

Figure 5: Schematic representation of the anatomical locations of five cortical point sources (see Table 2) used for generating EEG recordings. The effective causal connections in the form of red arrows correspond to the autoregressive model specified in Appendix 2.

Scalp electric potentials were computed at only 19 electrodes, corresponding to the 10/20 electrode placement system, in a realistic head model (Fuchs et al. 2002), using the MNI152 template (Mazziotta et al. 2001). The sLORETA (Pascual-Marqui 2002) inverse solution was computed at 6239 grey matter voxels at 5 mm spatial resolution. After estimating all 6239 signals, the inverse solution signals at the five cortical generator voxels were used for connectivity analysis.

Figure 6 shows the isolated effective coherence (iCoh) as a function of frequency (Pascual-Marqui et al 2014) for the actual signals, for the inverse solution signals, and for the "leakage corrected" inverse solution signals.

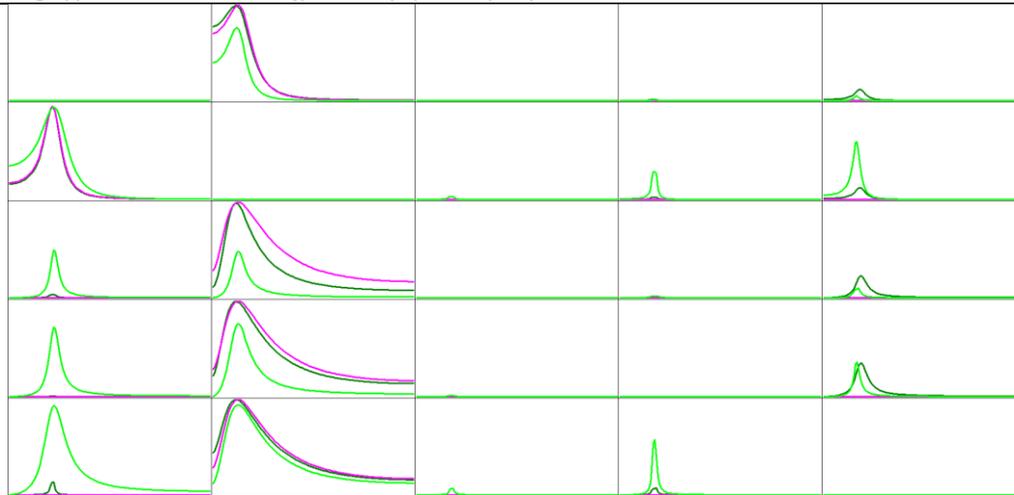

Figure 6: Isolated effective coherence (iCoh) as a function of frequency, for the toy example in Figure 1 and Appendix 2. The number of time samples used in the toy example is 25600. Horizontal frequency axis range 1-127 Hz. Vertical iCoh axis range 0-1. A column is a causal sender, a row is a causal receiver. Magenta curves are correct values, and dark green curves correspond to the inverse solution signals, as replicated from Pascual-Marqui et al 2014. Light green curves correspond to the "leakage-corrected" inverse solution signals, following the methodological prescription of Colclough et al 2015.

In Figure 6, a column is a causal sender, and a row is a causal receiver. Magenta curves are correct values, while dark green curves correspond to the inverse solution signals, as replicated from Pascual-Marqui et al 2014. Note that the dark green curves based on the estimated inverse signals give almost perfect connectivity results, matching the magenta curves. This means that the estimated inverse solution signals are very close to the true signals, which is as expected, since the generators are well separated.

The light green curves in Figure 6 correspond to the "leakage-corrected" of the inverse solution signals, following the methodological prescription of Colclough et al 2015.

As can be seen, leakage correction introduces many false connections when using estimated inverse solution signals.

## 9. Correct human connectomes ensue from innovations orthogonalization under very broad conditions, even with signal mixing

Here we use the same signals from section "5. False human connectomes ensue from leakage correction, i.e. signal orthogonalization, under very broad conditions, even without mixing", consisting of 5 signals from an AR model of order 2. In addition, analysis is performed on very strongly mixed signals, with a mixing matrix with off-diagonal elements equal to 0.7.

We use the procedure outlined in section "3. The resolution of a mixture of time series by innovations orthogonalization".

In this case, an AR model of order 2 we selected based on the AIC criterion, both for the original signals and for the strongly mixed signals. The estimated mixing matrices are very near the theoretical ones, as shown in Table 3.

Table 3: Estimated mixing matrices based on innovations orthogonalization (Eq. 15). Table 3a corresponds to the original unmixed five signals, which has a theoretical mixing matrix equal to the identity matrix. Table 3b corresponds to the strongly mixed signals, which has a theoretical mixing matrix with off-diagonal elements equal to 0.7

Table 3a

| 1.000 | -0.001 | -0.001 | 0.001 | 0.000 |
|---|---|---|---|---|
| -0.001 | 1.000 | -0.005 | 0.001 | 0.005 |
| -0.001 | -0.005 | 1.000 | 0.001 | -0.003 |
| 0.001 | 0.001 | 0.001 | 1.000 | 0.003 |
| 0.000 | 0.005 | -0.003 | 0.003 | 1.000 |

Table 3b

| 1.000 | 0.700 | 0.699 | 0.701 | 0.700 |
|---|---|---|---|---|
| 0.700 | 1.000 | 0.698 | 0.700 | 0.702 |
| 0.699 | 0.698 | 1.000 | 0.700 | 0.699 |
| 0.701 | 0.700 | 0.700 | 1.000 | 0.701 |
| 0.700 | 0.702 | 0.699 | 0.701 | 1.000 |

Using the method of innovations orthogonalization, Figure 7 shows the isolated effective coherence (iCoh) as a function of frequency (Pascual-Marqui et al 2014) for the actual signals, for the estimated unmixed signals based on the actual signals, and for the estimated unmixed signals based on the strongly mixed signals.

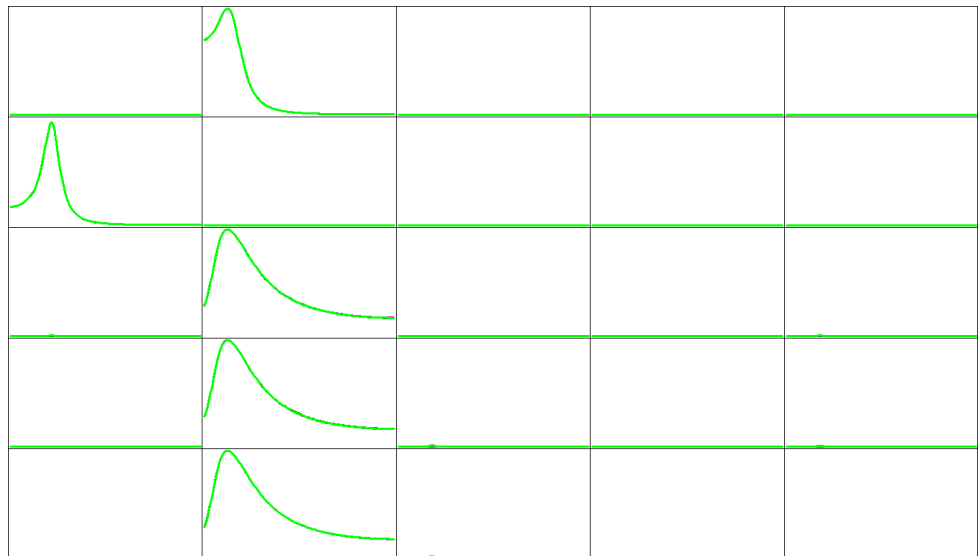

Figure 7: Isolated effective coherence (iCoh) as a function of frequency, for the toy example in Figure 1 and Appendix 2. The number of time samples used in the toy example is 25600. Horizontal frequency axis range 1-127 Hz. Vertical iCoh axis range 0-1. A column is a causal sender, a row is a causal receiver. Three results are shown, but only one is perceived because all results are almost identical: actual signals, innovations orthogonalization applied to the actual signals, and innovations orthogonalization applied to the strongly mixed signals.

This example shows that innovations orthogonalization is capable of correctly resolving mixtures of signals.

Two very important conclusions follow:
1. Innovations orthogonalization is capable of correctly identifying unmixed signals, even if they have significant non-zero values of cross-correlation at lag-zero.
2. Innovations orthogonalization is capable of correctly resolving (i.e. unmixing) strongly mixed signals.

## 10. Correct human connectomes ensue from innovations orthogonalization of signals of electric neuronal activity in oscillatory models, with and without mixing

Here we use the same signals from section "6. False human connectomes ensue from leakage correction, i.e. signal orthogonalization, of signals of electric neuronal activity in oscillatory models, even without mixing", consisting of 3 noisy oscillators.

We use the procedure outlined in section "3. The resolution of a mixture of time series by innovations orthogonalization".

In addition, we strongly mix the three signals with a mixing matrix having a value of 0.7 for all off-diagonal elements. In both cases, for the unmixed and strongly mixed signals, an AR model of order 9 was used for orthogonalizing the innovations. The choice of the AR order in this case is relatively arbitrary, because the AIC criterion monotonically decreases up to AR order of 20 (highest AR order tested was 20). This is expected, since these signals are not generated by an AR model. Nevertheless, an AR order of 9 suffices for a "good fit". The coherences for the original signals, and the resolved signals (unmixed and mixed), all overlap, giving essentially identical coherence values for all frequencies, as shown in the magenta curve in Figure 3.

Three very important conclusions follow:
1. Innovations orthogonalization is capable of correctly identifying unmixed signals, even if they have significant non-zero values of cross-correlation at lag-zero.
2. Innovations orthogonalization is capable of correctly resolving (i.e. unmixing) strongly mixed signals.
3. The two previous conclusions apply to stochastic signals that are sinusoidal in nature, even if they are not generated by a proper AR mechanism.

## 11. Correct human connectomes ensue from innovations orthogonalization of signals analyzed for amplitude-amplitude coupling analysis

In this case the method of innovations orthogonalization is applied to two sets of amplitude-modulated signals:
1. The three signals specified in the section "7. False human connectomes ensue from leakage correction, i.e. signal orthogonalization, in amplitude-amplitude coupling analysis", see Eq. 27, Eq. 28, Eq. 29, and Appendix 5.
2. The strongly mixed signals, using a mixing matrix with off-diagonal elements equal to 0.7.

Recall that these stochastic signals have independent instantaneous amplitudes (i.e. envelopes). This means that there is no amplitude-amplitude coupling, although the signals are amplitude modulated.

For the innovations orthogonalization, an AR model of order 9 was arbitrarily chosen, which was sufficiently high for achieving a good fit. The AIC criterion was monotonically decreasing up to AR order 20.

The estimated mixing matrices were near ideal, with maximum absolute deviation equal to 0.004 from the theoretical matrix.

After unmixing, the estimated unmixed signals were used for computing the instantaneous amplitude signals. For the unmixed signals, the maximum amplitude-amplitude correlation among all pairs of signals was -0.0085, and for the mixed signals it was -0.0083, which is not significant (sample size 25600).

Three very important conclusions follow:
1. Innovations orthogonalization is capable of correctly identifying unmixed signals, even if they have significant non-zero values of cross-correlation at lag-zero.
2. Innovations orthogonalization is capable of correctly resolving (i.e. unmixing) strongly mixed signals.
3. The two previous conclusions apply to stochastic signals that are amplitude modulated in nature, even if they are not generated by a proper AR mechanism.

## 12. Correct human connectomes ensue from innovations orthogonalization of inverse solution signals

The signals studied here follow closely the ones used above in section "8. False human connectomes ensue from leakage correction, i.e. signal orthogonalization, of inverse solution signals." This corresponds to five generators with cortical locations specified in Table 2, and with time course of activity generated by the AR model specified in Appendix 2.

But in this section, instead of analyzing estimated inverse solution signals at only the same five generator locations, we will analyze six signals, with the first five corresponding to the actual generator locations, and a sixth signal where there is no generator (illustrated in Figure 8):
`Superior frontal gyrus (medial); Brodmann area 6; (X= 5 , Y= -10 , Z= 70) (MNI coords mm)`

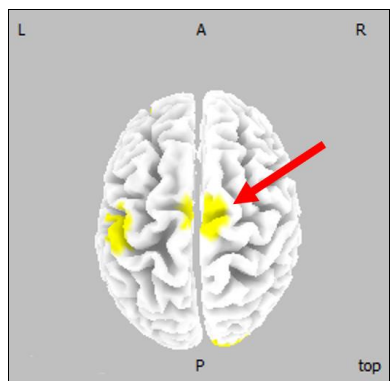

Figure 8: The red arrow indicates an additional 6th region of interest:
`Superior frontal gyrus (medial); Brodmann area 6; (X= 5 , Y= -10 , Z= 70) (MNI coords mm)`
where there is no generator, to be used in connectivity analysis of estimated inverse solution signals. The actual 5 generators are specified in Table 2 and displayed in Figure 5. Notation: L=left, R=right, A=anterior, P=posterior, top=top view.

The purpose of adding a region of interest without a generator is to expose an important failure of connectivity analysis of inverse solution signals.

Figure 9 displays the isolated effective coherence (iCoh) as a function of frequency (Pascual-Marqui et al 2014) for the six inverse solution signals, and for the unmixed signals using innovations orthogonalization, based on an AR model of order 2.

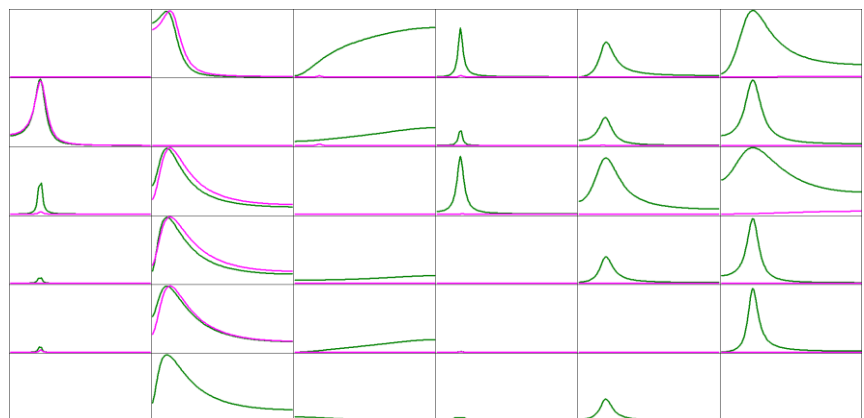

Figure 9: Isolated effective coherence (iCoh) as a function of frequency, for six inverse solution signals, with the first five corresponding to actual connected generators (see Table 2, Figure 5). The 6th region without actual generator is shown in Figure 8. Horizontal frequency axis range 1-127 Hz. Vertical iCoh axis range 0-1. A column is a causal sender, a row is a causal receiver. Magenta curves correspond to unmixed inverse solution signals using innovations orthogonalization. Dark green curves correspond to the estimated inverse solution signals without applying any unmixing procedure.

The green curves in Figure 9 correspond to iCoh measures for the six estimated inverse solution signals without applying any unmixing procedure. This result exposes a major pitfall of using inverse solution signals for connectivity analysis: the simple inclusion of the 6th region without generator gives an extremely distorted and false connectivity pattern as compared to the use of only the five generator regions (see dark green curves in Figure 6).

However, the magenta curves correspond to iCoh measures for the innovations orthogonalization method applied to the six inverse solution signals. The results give an almost perfect match to the theoretical values.

## 13. Why does the method of innovations orthogonalization work?

The method of innovations orthogonalization is based on the very powerful "innovations approach" to time series analysis. A major treatise on this methodology can be found in Ozaki 2012.

Intuitively, the assumption of independent innovations is a form of constraint on the dynamical inverse solution, which results in increased spatial resolution, i.e. "less mixing of the signals". An important example of spatio-temporal inverse solutions that display higher spatial resolution can be found in Galka et al 2004.

In this work, the assumed dynamical model is a simple AR model. Other forms can be accommodated, such as nonlinear AR models.

Obviously, the method will fail if the assumed dynamical model (linear, nonlinear, stationary, nonstationary, etc) has correlated innovations.

## 14. Related models

Erla et al 2009 and Hyvarinen et al 2010 considered a multivariate autoregressive model that starts at lag zero:

**Eq. 30** $$\mathbf{Z}_t = \sum_{k=0}^{q} \mathbf{C}_k \mathbf{Z}_{t-k} + \mathbf{e}_t$$

equivalent to:

**Eq. 31** $$(\mathbf{I} - \mathbf{C}_0)\mathbf{Z}_t = \sum_{k=1}^{q} \mathbf{C}_k \mathbf{Z}_{t-k} + \mathbf{e}_t$$

and:

**Eq. 32** $$\mathbf{Z}_t = \sum_{k=1}^{q} \left[ (\mathbf{I} - \mathbf{C}_0)^{-1} \mathbf{C}_k \right] \mathbf{Z}_{t-k} + \left[ (\mathbf{I} - \mathbf{C}_0)^{-1} \mathbf{e}_t \right]$$

where the matrix $\mathbf{C}_0$ contains the coefficients related to the instantaneous effect that one variable exerts on another.

We note the similarity in form of Eq. 32 with our present work in Eq. 1, Eq. 9, Eq. 10, and Eq. 11.

However, in our present work, we assume that there are no physiologically mediated instantaneous interactions between cortical regions, which means that we assume that $\mathbf{C}_0 = \mathbf{0}$. Therefore, the similarity between our model and the autoregressive model with instantaneous effects (Eq. 30, Eq. 31, Eq. 32) is only formal.

Another important difference between the models is that in order to estimate the instantaneous coefficients in $\mathbf{C}_0$, Hyvarinen et al 2010 assume non-Gaussianity for the independent innovations $\mathbf{e}_t$ and that the matrix $\mathbf{C}_0$ corresponds to an acyclic graph. Under these conditions, a straightforward adaptation of the powerful linear non-Gaussian acyclic model (LiNGAM) of Shimizu et al 2006 is used for the estimation.

## 16. Appendix 1: Orthogonal Procrustes Algorithm

The statement of this problem and the algorithm can be found in Everson 1999 and in Colclough et al 2015. Given a full rank matrix $\mathbf{A} \in \mathbb{R}^{N \times p}$ with $N > p$, the aim is to approximate it as:

**Eq. 33**    $\mathbf{A} \approx \mathbf{VD}$

where $\mathbf{V} \in \mathbb{R}^{N \times p}$ is orthonormal:

**Eq. 34**    $\mathbf{V}^T \mathbf{V} = \mathbf{I}$

and $\mathbf{D} \in \mathbb{R}^{p \times p}$ is diagonal:

**Eq. 35**    $\mathbf{D} = diag(\mathbf{D})$

This can be stated as a constrained minimization problem:

**Eq. 36**    $\min_{\mathbf{V},\mathbf{D}} tr\left[(\mathbf{A} - \mathbf{VD})^T (\mathbf{A} - \mathbf{VD})\right]$ , under constraints: $\mathbf{V}^T \mathbf{V} = \mathbf{I}$ , $\mathbf{D} = diag(\mathbf{D})$

The following algorithm solves this problem (its derivation appears in Everson 1999 and in Colclough et al 2015). Formally, input a matrix $\mathbf{A}$ and the algorithm returns matrices $\mathbf{V}$ and $\mathbf{D}$.

**Step-1**: Initialize the diagonal matrix $\mathbf{D} \in \mathbb{R}^{p \times p}$ to the identity matrix.
**Step-2**: Compute the SVD (singular value decomposition) of $(\mathbf{AD})$ as:

**Eq. 37**    $(\mathbf{AD}) = \mathbf{L} \mathbf{\Lambda} \mathbf{R}^T$

where $\mathbf{L} \in \mathbb{R}^{N \times p}$ contains the orthonormal left eigenvectors in the columns, $\mathbf{R} \in \mathbb{R}^{p \times p}$ contains the orthonormal right eigenvectors in the columns, and the diagonal matrix $\mathbf{\Gamma} \in \mathbb{R}^{p \times p}$ contains the eigenvalues.
**Step-3**: Compute the matrix $\mathbf{V} \in \mathbb{R}^{N \times p}$ as:

**Eq. 38**    $\mathbf{V} = \mathbf{L} \mathbf{R}^T$

**Step-4**: Compute the matrix $\Delta \in \mathbb{R}^{p \times p}$ as:

**Eq. 39**    $\Delta = \mathbf{V}^T \mathbf{A}$

**Step-5**: Compute the diagonal matrix $\mathbf{D}$ as:

**Eq. 40**    $\mathbf{D} = diag(\Delta)$

**Step-6**: Go to step 2 until convergence (e.g. until the matrix $\mathbf{D}$ changes negligibly from one iteration to the next).

## 17. Appendix 2: An autoregressive model of order=2, for five time series

An autoregressive model of order=2, for five time series, is:

$$\left[ \mathbf{X}_t = \mathbf{A}(1)\mathbf{X}_{t-1} + \mathbf{A}(2)\mathbf{X}_{t-2} + \boldsymbol{\varepsilon}_t \right]$$

with independent and identically distributed Gaussian innovations $\boldsymbol{\varepsilon} \sim \mathbf{N}(\mathbf{0}, \mathbf{I})$, and with coefficients:

A(1)=

| 1.5  | -0.25 | 0    | 0    | 0    |
|------|-------|------|------|------|
| -0.2 | 1.8   | 0    | 0    | 0    |
| 0    | 0.9   | 1.65 | 0    | 0    |
| 0    | 0.9   | 0    | 1.65 | 0    |
| 0    | 0.9   | 0    | 0    | 1.65 |

A(2)=

| -0.95 | 0     | 0     | 0     | 0     |
|-------|-------|-------|-------|-------|
| 0     | -0.96 | 0     | 0     | 0     |
| 0     | -0.8  | -0.95 | 0     | 0     |
| 0     | -0.8  | 0     | -0.95 | 0     |
| 0     | -0.8  | 0     | 0     | -0.95 |

## 18. Appendix 3: Connections in a 5-node causal network, with and without leakage correction, with and without mixing the signals

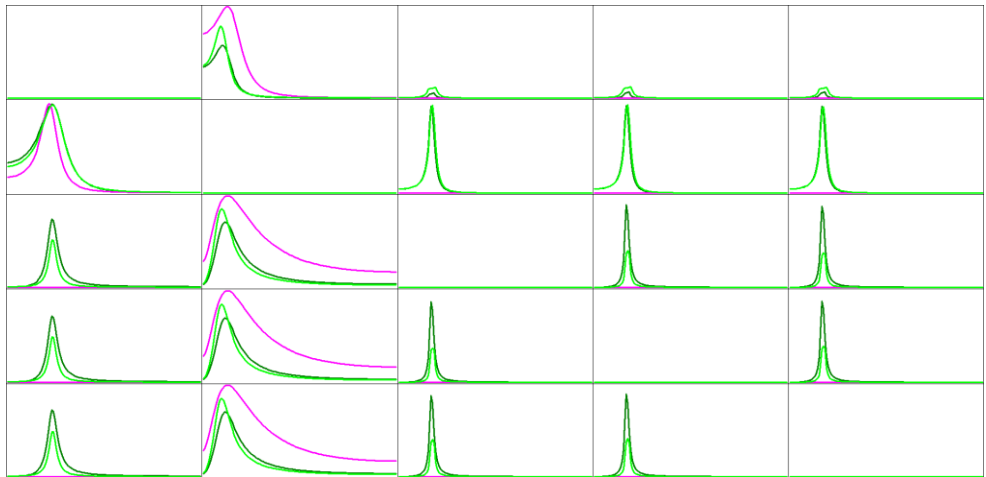

Figure A3-1: Isolated effective coherence (iCoh) as a function of frequency, for the toy example in Figure 1 and Appendix 2. The number of time samples used in the toy example is 25600. Horizontal frequency axis range 1-127 Hz. Vertical iCoh axis range 0-1. A column is a causal sender, a row is a causal receiver. Magenta curves are correct values, and dark green curves are for the "leakage-corrected" signal, corresponding to Figure 2 in the main text. The additional light green curves correspond to leakage correction applied to the mixed signals with instantaneous mixing matrix having ones on the diagonal and 0.7 on all off-diagonal elements.

# 19. Appendix 4: Code for generating noisy oscillatory activity

PASCAL-type code for generating noisy oscillatory activity

```
var ep,t1,t,nep,ntt,nsig:integer; x:matrix; w1,w2,w3,d1,d2,d3:real;
    a1,a2,a3,u1,u2:real;
begin
  nep:=100;
  ntt:=256;
  nsig:=3;
  w1:=10; w2:=10; w3:=17;
  d1:=0; d2:=-1; d3:=-2;
  u1:=0.5; u2:=0.9;

  randomize;
  t:=0;
  create1(nep*ntt,nsig,x);
  for ep:=1 to nep do begin
    a1:=urand.Random1(u1);
    a2:=urand.Random1(u1);
    a3:=urand.Random1(u1);
    for t1:=1 to ntt do begin
      inc(t);
      x.m[t,kk1]:=a1*sin(2*pi*(t-d1)*w1/ntt)+urand.random0(u2);
      x.m[t,kk2]:=a2*sin(2*pi*(t-d2)*w2/ntt)+urand.random0(u2);
      x.m[t,kk3]:=a3*sin(2*pi*(t-d3)*w3/ntt)+urand.random0(u2);
    end;
  end;
  writetxt1('!3SineWaves.txt',x);
  destroy1(x);
  showmessage('dun');
  halt;
end;

// urand.Random1(u) returns uniform rand in interval 1-u1 to 1+u1
// urand.Random0(u2) returns uniform rand in interval -u2 to +u2
```

## 20. Appendix 5: Code for generating amplitude modulated signals

PASCAL-type code for generating amplitude modulated signals

```
var nt,deltat,t:integer; a,sr,ff1,ff2,ff3,sd,d1,d2,d3,sf1,sf2,sf3:real;
    InstAmp,AmpMod:matrix;
begin
  nt:=25600; deltat:=4;
  a:=0.5; sr:=256; ff1:=22; ff2:=22; ff3:=28; sd:=0.2;
  d1:=0; d2:=-1; d3:=-2;
  sf1:=2; sf2:=3; sf3:=5;
  create1(nt,3,instamp);
  create1(nt,3,ampmod);
  fillchar1(instamp,0);
  fillchar1(ampmod,0);

  for t:=1 to nt do begin
    instamp.m[t,kk1]:=(a+(1+sin(2*pi*sf1*(t)/sr))/2)*random1(sd);
    instamp.m[t,kk2]:=(a+(1+sin(2*pi*sf2*(t-deltat)/sr))/2)*random1(sd);
    instamp.m[t,kk3]:=(a+(1-sin(2*pi*sf3*(t-deltat)/sr))/2)*random1(sd);
  end;

  for t:=1 to nt do begin
    ampmod.m[t,kk1]:=instamp.m[t,kk1]*sin(2*pi*ff1*(t-d1)/sr);
    ampmod.m[t,kk2]:=instamp.m[t,kk2]*sin(2*pi*ff2*(t-d2)/sr);
    ampmod.m[t,kk3]:=instamp.m[t,kk3]*sin(2*pi*ff3*(t-d3)/sr);
  end;

  writetxt1('!InstAmp3Sigs.txt',instamp);
  writetxt1('!AmpMod3SigsSM00.txt',ampmod);

  SmoothSignals(0.7,ampmod);
  writetxt1('!AmpMod3SigsSM70.txt',ampmod);

  destroy1(instamp);
  destroy1(ampmod);
end;
```

## 21. Acknowledgments

This work was partially supported by the Center of Innovation (COI) Program, under the Japan Science and Technology Agency (JST).